\documentclass{article}
\usepackage{spconf,amsmath,epsfig}
\usepackage{amsfonts}
\usepackage[latin1]{inputenc}
\usepackage{amssymb}
\usepackage{latexsym}
\usepackage{amsfonts}
\usepackage{amsbsy}
\usepackage{amsmath}
\usepackage{times}
\usepackage{epsfig,epsf,times,amssymb,amsmath}

\title{Adaptive Low-rank Constrained Constant Modulus Beamforming Algorithms using Joint Iterative Optimization of Parameters}

\name{Lei Wang and Rodrigo C. de Lamare}
\address{Communications Research Group, Department of Electronics, University of York, YO10 5DD, UK.\\Email:\{lw517,rcdl500\}@ohm.york.ac.uk}

\begin{document}
\ninept \maketitle \linespread{0.9}
\begin{abstract}
This paper proposes a robust reduced-rank scheme for adaptive
beamforming based on joint iterative optimization (JIO) of adaptive
filters. The scheme provides an efficient way to deal with filters
with large number of elements. It consists of a bank of full-rank
adaptive filters that forms a transformation matrix and an adaptive
reduced-rank filter that operates at the output of the bank of
filters. The transformation matrix projects the received vector onto
a low-dimension vector, which is processed by the reduced-rank
filter to estimate the desired signal. The expressions of the
transformation matrix and the reduced-rank weight vector are derived
according to the constrained constant modulus (CCM) criterion. Two
novel low-complexity adaptive algorithms are devised for the
implementation of the proposed scheme with respect to different
constrained conditions. Simulations are performed to show superior
performance of the proposed algorithms in comparison with the
existing methods.

\end{abstract}
\section{Introduction}

Adaptive beamforming technology is of paramount importance in
numerous signal processing applications such as radar, wireless
communications, and sonar \cite{Litva}-\cite{Jian}. Among various
beamforming techniques, the beamformers according to the constrained
minimum variance (CMV) criterion \cite{Jian} are prevalent, which
minimize the total output power while maintaining the gain along the
direction of the signal of interest (SOI). Another alternative
beamformer design is performed according to the constrained constant
modulus (CCM) criterion, which is a positive measure \cite{Jian} of
the average amount that the beamformer output deviates from a
constant modulus condition. Compared with the CMV, the CCM
beamformers exhibit superior performance in many severe scenarios
(e.g., steering vector mismatch) since the positive measure provides
more information for the parameter estimation.

Many adaptive algorithms have been developed according to the CMV
and CCM criteria for implementation. A simple and popular one is
stochastic gradient (SG) method \cite{Frost}, \cite{Haykin}. A major
drawback of the SG-based methods is that, when the number of
elements in the filter is large, they always require a large amount
of samples to reach the steady-state. Furthermore, in dynamic
scenarios, filters with many elements usually show poor performance
in tracking signals embedded in interference and noise. Reduced-rank
signal processing was motivated to provide a way out of this dilemma
\cite{Guerci}, \cite{Honig1}. For the application of beamforming,
the reduced-rank technique project the received vector onto a
low-dimension subspace and perform the filter optimization within
this subspace. One popular reduced-rank scheme is the multistage
wiener filter (MSWF), which employs the minimum mean squared error
(MMSE) \cite{Goldstein}, and its extended versions that utilize the
CMV and CCM criteria were reported in \cite{Honig}, \cite{Lamare}.
Another technique that resembles the MSWF is the auxiliary-vector
filtering (AVF) \cite{Pados}, \cite{Pados2}. Despite improved
convergence and tracking performance achieved by these methods,
their implementations require high computational cost and suffer
from numerical problems. A joint iterative optimization (JIO)
scheme, which was presented recently in \cite{Lamare1}, employs the
CMV criterion with a low-complexity adaptive implementation to
achieve better performance than the existing methods.

Considering the fact that the CCM-based beamformers outperform the
CMV ones for constant modulus constellations, we propose a robust
reduced-rank scheme according to the CCM criterion for the
beamformer design. The proposed reduced-rank scheme consists of a
bank of full-rank adaptive filters, which constitutes the
transformation matrix, and an adaptive reduced-rank filter that
operates at the output of the bank of filters. The transformation
matrix projects the full-rank received vector onto a low-dimension,
which is then processed by the reduced-rank filter to estimate the
desired signal. The transformation matrix and the reduced-rank
filter are computed based on the JIO. The proposed scheme provides
an iterative exchange of information between the transformation
matrix and the reduced-rank filter, which leads to improved
convergence and tracking ability and low-complexity cost. We devise
two adaptive algorithms for the implementation of the proposed
reduced-rank scheme. The first one employs the SG approach to
jointly estimate the transformation matrix and the reduced-rank
weight vector subject to a constraint on the array response. The
second proposed algorithm is extended from the first one and
reformulates the transformation matrix subject to an orthogonal
constraint. The Gram Schmidt (GS) technique \cite{Golub} is employed
to realize the reformulation. The performance of the second method
outperforms the first one. Simulation results are given to
demonstrate the preferable performance and stability achieved by the
proposed algorithms versus the existing methods in typical
scenarios.

\section{System Model and Problem Statement}

\subsection{System Model}

Let us suppose that $q$ narrowband signals impinge on an uniform
linear array (ULA) of $m$ ($m\geq q$) sensor elements. The sources
are assumed to be in the far field with directions of arrival (DOAs)
$\theta_{0}$,\ldots,$\theta_{q-1}$. The $i$th received vector
$\boldsymbol x(i)\in\mathbb C^{m\times 1}$ can be modeled as
\begin{equation} \label{1}
\centering {\boldsymbol x}(i)={\boldsymbol {A}}({\boldsymbol
{\theta}}){\boldsymbol s}(i)+{\boldsymbol n}(i),~~~ i=1,\ldots,N
\end{equation}
where
$\boldsymbol{\theta}=[\theta_{0},\ldots,\theta_{q-1}]^{T}\in\mathbb{C}^{q
\times 1}$ is the signal DOAs, ${\boldsymbol A}({\boldsymbol
{\theta}})=[{\boldsymbol a}(\theta_{0}),\ldots,{\boldsymbol
a}(\theta_{q-1})]\in\mathbb{C}^{m \times q}$ comprises the signal
direction vectors ${\boldsymbol a}(\theta_{k})=[1,e^{-2\pi
j\frac{d}{\lambda_{c}}cos{\theta_{k}}},\ldots$, $e^{-2\pi
j(m-1)\frac{d}{\lambda_{c}}cos{\theta_{k}}}]^{T}\in\mathbb{C}^{m
\times 1}$, $(k=0,\ldots,q-1)$, where $\lambda_{c}$ is the
wavelength and $d$ is the inter-element distance of the ULA
($d=\lambda_c/2$ in general), and to avoid mathematical ambiguities,
the direction vectors $\boldsymbol a(\theta_{k})$ are considered to
be linearly independents. ${\boldsymbol s}(i)\in \mathbb C^{q\times
1}$ is the source data, ${\boldsymbol n}(i)\in\mathbb C^{m\times 1}$
is temporarily white sensor noise, which is assumed to be a
zero-mean spatially and Gaussian process, $N$ is the observation
size of snapshots, and $(\cdot)^{T}$\ stands for transpose. The
estimated desired signal is given by
\begin{equation} \label{2}
\centering y(i)={\boldsymbol w}^H(i) {\boldsymbol x}(i)
\end{equation}
where ${\boldsymbol w}(i)=[w_{1}(i),\ldots,w_{m}(i)]^{T}\in\mathbb
C^{m\times 1}$ is the complex weight vector, and $(\cdot)^{H}$
stands for Hermitian transpose.

\subsection{Problem Statement}

Let us consider the full-rank CCM filter for beamforming, which can
be computed by minimizing the following cost function
\begin{equation}\label{3}
J_{\textrm{cm}}\big(\boldsymbol w(i)\big)=\mathbb
E\big\{\big[|y(i)|^{2}-1\big]^{2}\big\},~~
\textrm{subject~to}~~{\boldsymbol w}^{H}(i){\boldsymbol
a}(\theta_{0})=1
\end{equation}
where $\theta_0$ is the direction of the SOI and $\boldsymbol
a(\theta_{0})$ denotes the corresponding normalized steering vector.
The cost function is the expected deviation of the squared modulus
of the array output to a constant subject to the constraint on the
array response, which is set to capture the power of the desired
signal and ensure the convexity of the cost function. The weight
expression obtained from (\ref{3}) is
\begin{equation}\label{4}
\boldsymbol w(i+1)=\boldsymbol R^{-1}(i)\Big\{\boldsymbol
p(i)-\frac{\big[\boldsymbol p^H(i)\boldsymbol R^{-1}(i)\boldsymbol
a(\theta_0)-1\big]\boldsymbol a(\theta_0)}{\boldsymbol
a^H(\theta_0)\boldsymbol R^{-1}(i)\boldsymbol a(\theta_0)}\Big\}
\end{equation}
where $\boldsymbol R(i)=\mathbb E[|y(i)|^2\boldsymbol
x(i)\boldsymbol x^H(i)]$, $\boldsymbol p(i)=\mathbb
E[y^{\ast}(i)\boldsymbol x(i)]$, and $(\cdot)^{\ast}$ denotes
complex conjugate. Note that (\ref{4}) is a function of previous
values of $\boldsymbol w(i)$ (since $y(i)=\boldsymbol
w^H(i)\boldsymbol x(i)$) and thus must be initialized to start the
iteration. We keep the time index in $\boldsymbol R(i)$ and
$\boldsymbol p(i)$ for the same reason. It is obvious that the
calculation of weight vector requires high complexity due to the
matrix inversion. The SG type algorithms can be employed to reduce
the computational load but still suffer from slow convergence and
tracking performance when the dimension $m$ is large. The
reduced-rank schemes like MSWF and AVF can be used to improve the
performance but still need high computational cost and suffer from
numerical problems.

\section{Proposed Reduced-rank Scheme and CCM Filters Design}
In this section, by proposing a reduced-rank scheme based on the JIO
of adaptive filters, we introduce a minimization problem according
to the CM criterion subject to different constraints. The
reduced-rank CCM filters design is described in details.

\subsection{Proposed Reduced-Rank Scheme}
Define a transformation matrix $\boldsymbol T_{r}(i)=[\boldsymbol
t_1(i), \boldsymbol t_2(i), \ldots,$ $\boldsymbol t_r(i)]\in\mathbb
C^{m\times r}$, which is responsible for the dimensionality
reduction, to project the $m\times 1$ received vector $\boldsymbol
x(i)$ onto a lower dimension, yielding
\begin{equation}\label{5}
\bar{\boldsymbol x}(i)=\boldsymbol T_{r}^{H}(i)\boldsymbol x(i)
\end{equation}
where $\boldsymbol t_l(i)=[t_{1,l}(i), \ldots,
t_{m,l}(i)]^T\in\mathbb C^{m\times 1},~l=1, \ldots, r$, makes up the
transformation matrix $\boldsymbol T_r(i)$, $\bar{\boldsymbol
x}(i)\in\mathbb C^{r\times 1}$ is the projected received vector, and
in what follows, all $r$-dimensional quantities are denoted by an
over bar. Here, $r<m$ is the rank and, as we will see, impacts the
output performance. An adaptive reduced-rank filter represented by
$\bar{\boldsymbol
w}(i)=[\bar{w}_{1}(i),\ldots,\bar{w}_{r}(i)]^{T}\in\mathbb
C^{r\times 1}$ is followed to get the filter output
\begin{equation}\label{6}
y(i)=\bar{\boldsymbol w}^H(i)\boldsymbol T_r^{H}(i)\boldsymbol x(i)
\end{equation}

\begin{figure}[htb]
\begin{minipage}[h]{1.0\linewidth}
  \centering
  \centerline{\epsfig{figure=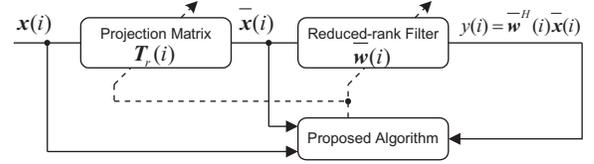,scale=0.7}} \vspace{-1em}\caption{Proposed reduced-rank beamforming scheme.} \label{fig:model1}
\end{minipage}
\end{figure}

From (\ref{6}), the filter output $y(i)$ depends on $\boldsymbol
T_r(i)$ and $\bar{\boldsymbol w}(i)$, as shown in Fig.
\ref{fig:model1}. It is necessary to jointly estimate $\boldsymbol
T_r(i)$ and $\bar{\boldsymbol w}(i)$ to get $y(i)$. We consider a
reduced-rank CM minimization problem subject to different
constraints, which are
\begin{equation}\label{7}
\begin{split}
&J_{\textrm{cm,JIO}}\big(\boldsymbol T_r(i), \bar{\boldsymbol
w}(i)\big)=\mathbb E\big\{\big[|y(i)|^{2}-1\big]^{2}\big\}\\
&\textrm{subject~to}~\bar{\boldsymbol w}^H(i)\boldsymbol
T_r^H(i)\boldsymbol a(\theta_0)=1
\end{split}
\end{equation}
\begin{equation}\label{8}
\begin{split}
&J_{\textrm{cm,JIO}}\big(\boldsymbol T_r(i), \bar{\boldsymbol
w}(i)\big)=\mathbb E\big\{\big[|y(i)|^{2}-1\big]^{2}\big\}\\
&\textrm{subject~to}~\bar{\boldsymbol w}^H(i)\boldsymbol
T_r^H(i)\boldsymbol a(\theta_0)=1~~\textrm{and}~~\boldsymbol
T_r^H(i)\boldsymbol T_r(i)=\boldsymbol I
\end{split}
\end{equation}

Compared with (\ref{7}), the constraint in (\ref{8}) includes one
orthogonal constraint on the transformation matrix, which is to
reformulate $\boldsymbol T_r(i)$ for the performance improvement. In
the following part, we will derive the CCM expressions of
$\boldsymbol T_r(i)$ and $\bar{\boldsymbol w}(i)$ with respect to
(\ref{7}). The proposed adaptive algorithm for the implementation of
(\ref{7}) and the extended algorithm with respect to (\ref{8}) will
represent in Section 4.

\subsection{Design of Reduced-rank CCM Filters}
Substituting (\ref{6}) into (\ref{7}), the cost function can be
transformed by the method of Lagrange multipliers into an
unconstrained one, which is
\begin{equation}\label{9}
\begin{split}
J_{\textrm{un}}\big(\boldsymbol T_r(i), \bar{\boldsymbol
w}(i)\big)&=\mathbb E\Big\{\big[|\bar{\boldsymbol w}^H(i)\boldsymbol
T_r^H(i)\boldsymbol
x(i)|^2-1\big]^2\Big\}\\
&+2\mathfrak{R}\Big\{\lambda\big[\bar{\boldsymbol w}^H(i)\boldsymbol
T_r^H(i)\boldsymbol a(\theta_0)-1\big]\Big\}
\end{split}
\end{equation}
where $\lambda$ is a scalar Lagrange multiplier and the operator
$\mathfrak{R}(\cdot)$ selects the real part of the argument.
Assuming $\bar{\boldsymbol w}(i)$ is known, minimizing (\ref{9})
with respect to $\boldsymbol T_r(i)$ equal to a null matrix and
solving for $\lambda$, we have
\begin{equation}\label{10}
\begin{split}
&\boldsymbol T_r(i+1)=\boldsymbol R^{-1}(i)\Big\{\boldsymbol
p(i)\bar{\boldsymbol w}^H(i)-\\
&\frac{\big[\bar{\boldsymbol w}^H(i)\bar{\boldsymbol
R}_{\bar{w}}^{-1}(i)\bar{\boldsymbol w}(i)\boldsymbol
p^H(i)\boldsymbol R^{-1}(i)\boldsymbol a(\theta_0)-1\big]\boldsymbol
a(\theta_0)\bar{\boldsymbol w}^H(i)}{\bar{\boldsymbol
w}^H(i)\bar{\boldsymbol R}_{\bar{w}}^{-1}(i)\bar{\boldsymbol
w}(i)\boldsymbol a^H(\theta_0)\boldsymbol R^{-1}(i)\boldsymbol
a(\theta_0)}\Big\}\bar{\boldsymbol R}_{\bar{w}}^{-1}(i)
\end{split}
\end{equation}
where $\boldsymbol p(i)=\mathbb E[y^{\ast}(i)\boldsymbol
x(i)]\in\mathbb C^{m\times1}$, $\boldsymbol R(i)=\mathbb
E[|y(i)|^2\boldsymbol x(i)\boldsymbol x^H(i)]\in\mathbb C^{m\times
m}$, and $\bar{\boldsymbol R}_{\bar{w}}(i)=\mathbb
E[\bar{\boldsymbol w}(i)\bar{\boldsymbol w}^H(i)]\in\mathbb
C^{r\times r}$. Note that the reduced-rank weight vector
$\bar{\boldsymbol w}(i)$ depends on the received vectors that are
random in practice, thus $\bar{\boldsymbol R}_{\bar{w}}(i)$ is
full-rank and invertible. $\boldsymbol R(i)$ and $\boldsymbol p(i)$
are functions of previous values of $\boldsymbol T_r$ and
$\bar{\boldsymbol w}$ due to the presence of $y(i)$. Therefore, it
is necessary to initialize $\boldsymbol T_r(i)$ and
$\bar{\boldsymbol w}(i)$ to estimate $\boldsymbol R(i)$ and
$\boldsymbol p(i)$, and start the iteration.

On the other hand, assuming $\boldsymbol T_r(i)$ is known,
minimizing (\ref{9}) with respect to $\bar{\boldsymbol w}(i)$ equal
to a null vector and solving for $\lambda$, we obtain
\begin{equation}\label{11}
\bar{\boldsymbol w}(i+1)=\bar{\boldsymbol
R}^{-1}(i)\Big\{\bar{\boldsymbol p}(i)-\frac{\big[\bar{\boldsymbol
p}^H(i)\bar{\boldsymbol R}^{-1}(i)\bar{\boldsymbol
a}(\theta_0)-1\big]\bar{\boldsymbol a}(\theta_0)}{\bar{\boldsymbol
a}^H(\theta_0)\bar{\boldsymbol R}^{-1}(i)\bar{\boldsymbol
a}(\theta_0)}\Big\}
\end{equation}
where $\bar{\boldsymbol R}(i)=\mathbb E[|y(i)|^2\boldsymbol
T_r^H(i)\boldsymbol x(i)\boldsymbol x^H(i)\boldsymbol
T_r(i)]\in\mathbb C^{r\times r}$, $\bar{\boldsymbol p}(i)=\mathbb
E[y^{\ast}(i)\boldsymbol T_r^H(i)\boldsymbol x(i)]\in\mathbb
C^{r\times 1}$, and $\bar{\boldsymbol a}(\theta_0)=\boldsymbol
T_r^H(i)\boldsymbol a(\theta_0)\in\mathbb C^{r\times1}$.

The expressions in (\ref{10}) for the transformation matrix and
(\ref{11}) for the reduced-rank weight vector depend on each other
and so are not closed-form solutions. It is necessary to iterate
$\boldsymbol T_r$ and $\bar{\boldsymbol w}$ with initial values for
implementation. Therefore, the initialization is not only for
estimating $y$ but starting the iteration. The proposed scheme
provides an iterative exchange of information between the
transformation matrix and the reduced-rank filter, which leads to
improved convergence and tracking performance. They are jointly
estimated to solve the CCM minimization problem.

\section{Development of Adaptive Algorithms}
\subsection{Proposed Adaptive SG Algorithm for (\ref{7})}
We describe a simple adaptive algorithm for implementation of the
proposed reduced-rank scheme according to the minimization problem
in (\ref{7}). Assuming $\bar{\boldsymbol w}(i)$ and $\boldsymbol
T_r(i)$ are known, respectively, taking the instantaneous gradient
of (\ref{9}) with respect to $\boldsymbol T_r(i)$ and
$\bar{\boldsymbol w}(i)$, and setting them equal to null matrices,
we obtain
\begin{equation}\label{12}
\nabla\mathcal J_{T_r}=2e(i)y^{*}(i)\boldsymbol x(i)\bar{\boldsymbol
w}^H(i)+\lambda_{T_r}\boldsymbol a(\theta_{0})\bar{\boldsymbol
w}^H(i)
\end{equation}
\begin{equation}\label{13}
\nabla\mathcal J_{\bar{w}}=2e(i)y^{*}(i)\boldsymbol
T_r^H(i)\boldsymbol x(i)+\lambda_{\bar{w}}\boldsymbol
T_r^H(i)\boldsymbol a(\theta_{0})
\end{equation}
where $e(i)=|y(i)|^2-1$, $\lambda_{T_r}$ and $\lambda_{\bar{w}}$ are
the corresponding Lagrange multipliers. Following the gradient rules
$\boldsymbol T_r(i+1)=\boldsymbol T_r(i)-\mu_{T_r}\nabla\mathcal
J_{T_r}$ and $\bar{\boldsymbol w}(i+1)=\bar{\boldsymbol
w}(i)-\mu_{\bar{w}}\nabla\mathcal J_{\bar{w}}$, substituting
(\ref{12}) and (\ref{13}) into them, respectively, and solving
$\lambda_{T_r}$ and $\lambda_{\bar{w}}$ by employing the constraint
in (\ref{7}), we obtain the iterative solutions in the form
\begin{equation}\label{14}
\begin{split}
\boldsymbol T_r(i+1)=\boldsymbol T_r(i)-&\mu_{T_r}e(i)y^*(i)\big[\boldsymbol x(i)\bar{\boldsymbol w}^H(i)\\
&-\boldsymbol a(\theta_{0})\bar{\boldsymbol w}^H(i)\boldsymbol
a^H(i)\boldsymbol x(i)\big]
\end{split}
\end{equation}
\begin{equation}\label{15}
\bar{\boldsymbol w}(i+1)=\bar{\boldsymbol
w}(i)-\mu_{\bar{w}}e(i)y^*(i)\big[\boldsymbol
I-\frac{\bar{\boldsymbol a}(\theta_{0})\bar{\boldsymbol
a}^H(\theta_{0})}{\bar{\boldsymbol a}^H(\theta_{0})\bar{\boldsymbol
a}(\theta_{0})}\big]\bar{\boldsymbol x}(i)
\end{equation}
where $\mu_{T_r}$ and $\mu_{\bar{w}}$ are the corresponding step
sizes, which are small positive values. The transformation matrix
$\boldsymbol T_r(i+1)$ and the reduced-rank weight vector
$\bar{\boldsymbol w}(i+1)$ are jointly updated. The filter output
$y(i)$ is estimated after each iterative procedure with respect to
the CCM criterion. We denominate this algorithm as JIO-CCM.

\subsection{Extended Algorithm for (\ref{8})}
Now, we consider the minimization problem in (\ref{8}). As explained
before, the constraint is added to orthogonalize a set of vectors
$\boldsymbol t_l(i+1)$ for the performance improvement. We employ
the Gram-Schmidt (GS) technique \cite{Golub} to realize this
constraint. Specifically, the adaptive SG algorithm in (\ref{14}) is
implemented to obtain $\boldsymbol T_r(i+1)$. Then, the GS process
is performed to reformulate the transformation matrix, which is
\cite{Golub}
\begin{equation}\label{16}
\boldsymbol t_{l,\textrm{ort}}(i+1)=\boldsymbol
t_{l}(i+1)-\sum_{j=1}^{l-1}\textrm{proj}_{\boldsymbol
t_{j,\textrm{ort}}(i+1)}\boldsymbol t_l(i+1)
\end{equation}
where $\boldsymbol t_{l,\textrm{ort}}(i+1)$ is the normalized
orthogonal vector after GS process and $\textrm{proj}_{\boldsymbol
t_{j,\textrm{ort}}(i+1)}\boldsymbol t_l(i+1)={\boldsymbol
t_{j,\textrm{ort}}^H(i+1)}\boldsymbol t_l(i+1)\frac{{\boldsymbol
t_{j,\textrm{ort}}}(i+1)}{\boldsymbol
t_{j,\textrm{ort}}^H(i+1)\boldsymbol t_{j,\textrm{ort}}(i+1)}$ is a
reformulation operator.

The reformulated transformation matrix $\boldsymbol
T_{r,\textrm{ort}}(i+1)$ is constructed after we obtain a set of
orthogonal $\boldsymbol t_{l, \textrm{ort}}(i+1),~l=1, \ldots, r$.
By employing $\boldsymbol T_{r,\textrm{ort}}(i+1)$ to get
$\bar{\boldsymbol x}(i)$, $\bar{\boldsymbol a}(\theta_0)$, and
jointly update with $\bar{\boldsymbol w}(i+1)$ in (\ref{15}), the
performance can be further improved. Simulation results will be
given to show this result. We denominate this GS version algorithm
as JIO-CCM-GS, which is performed by computing (\ref{14}),
(\ref{16}), and (\ref{15}).

The computational complexity with respect to the existing and
proposed algorithms is evaluated according to additions and
multiplications. The complexity comparison is listed in Table
\ref{tab: Computational complexity}. The complexity of the proposed
JIO-CCM and JIO-CCM-GS algorithms increases with the multiplication
of $rm$. The parameter $m$ is more influential since $r$ is selected
around a small range that is much less than $m$ for large arrays,
which will be shown in simulations. This complexity is about $r$
times higher than the full-rank algorithms \cite{Frost}, slightly
higher than the recent JIO-CMV algorithm \cite{Lamare}, but much
lower than the MSWF-based \cite{Honig}, \cite{Lamare}, and AVF
\cite{Pados} methods.
\begin{table}
\centering \caption{\normalsize Computational complexity}
\footnotesize \label{tab: Computational complexity}
\begin{tabular}{l c c}
\hline \\
Algorithm & Additions                   & Multiplications\\
\hline   \\
Full-Rank-CMV    & $3m-1$                        & $4m+1$ \\
Full-Rank-CCM    & $3m$                  & $4m+3$ \\
MSWF-CMV         & $rm^2+rm+m$             & $rm^{2}+m^2+2rm$ \\
                 & $+2r-2$                            & $+5r+2$\\
MSWF-CCM        & $rm^2+rm+m$             & $rm^2+m^2+2rm$ \\
                & $+2r-1$                 & $+5r+4$\\
AVF           & $r(4m^2+m-2)$   & $r(5m^2+3m)$\\
              & $+5m^2-m-1$     & $+8m^2+2m$\\
JIO-CMV        & $4rm+m+2r-3$            & $4rm+m+7r+3$\\
JIO-CMV-GS     & $7rm-m-1$               & $7rm-2m+8r+2$\\
JIO-CCM        & $4rm+m+2r-2$             & $4rm+m+7r+6$\\
JIO-CCM-GS     & $7rm-m$                  & $7rm-2m+8r+5$\\
\hline
\end{tabular}
\end{table}

\section{Simulations}

Simulations are performed by an ULA containing $m=32$ sensor
elements with half-wavelength interelement spacing. We compare the
proposed JIO-CCM and JIO-CCM-GS algorithms with the full-rank
\cite{Frost}, MSWF \cite{Honig}, \cite{Lamare}, and AVF \cite{Pados}
methods and in each method, the CMV and CCM criteria are considered
with the SG algorithm for implementation. A total of $K=1000$ runs
are used to get the curves. In all experiments, the BPSK source
power (including the desired user and interferers) is
$\sigma_{s}^{2}=\sigma_{i}^2=1$ and the input SNR $=10$ dB with
spatially and temporally white Gaussian noise.

\begin{figure}[htb]
\begin{minipage}[h]{1.0\linewidth}
  \centering
  \centerline{\epsfig{figure=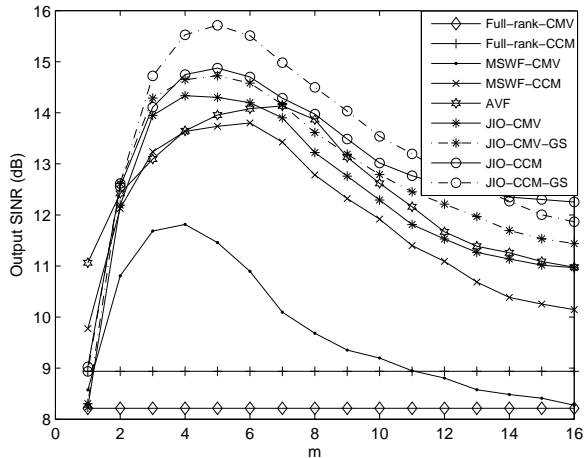,scale=0.6}} \vspace{-1.5em}\caption{Output SINR versus the
number of snapshots with $\mu_{T_r}=0.002$, $\mu_{\bar{w}}=0.001$,
$\mu_{T_r,\textrm{gs}}=0.003$, $\mu_{\bar{w},\textrm{gs}}=0.0007$.}
\label{fig:cmv_ccm_sg_gram_final}
\end{minipage}
\end{figure}

In Fig. \ref{fig:cmv_ccm_sg_gram_final}, we consider the presence of
$q=7$ users (one desired) in the system. The transformation matrix
and the reduced-rank weight vector are initialized with $\boldsymbol
T_r(0)=[\boldsymbol I_r^T~\boldsymbol 0_{r\times(m-r)}^T]$ and
$\bar{\boldsymbol w}(0)=\big(\boldsymbol T_{r}^{H}(0)\boldsymbol
a(\theta_{0})\big)/\big(\|\boldsymbol T_{r}^{H}(0)\boldsymbol
a(\theta_{0})\|^{2}\big)$ to ensure the constraint in (\ref{7}). The
rank is $r=r_{\textrm gs}=5$  for the proposed JIO-CCM and
JIO-CCM-GS algorithms. Fig. \ref{fig:cmv_ccm_sg_gram_final} shows
that all output SINR values increase to the steady-state as the
increase of the snapshots. The JIO-based algorithms have superior
steady-state performance as compared with the full-rank, MSWF, and
AVF methods. The GS version algorithms enjoy further developed
performance comparing with corresponding JIO-CMV and JIO-CCM
methods. Checking the convergence, the proposed algorithms are
slightly slower than the AVF, which is least squares (LS)-based, and
much faster than the other methods.

\begin{figure}[htb]
\begin{minipage}[h]{1.0\linewidth}
  \centering
  \centerline{\epsfig{figure=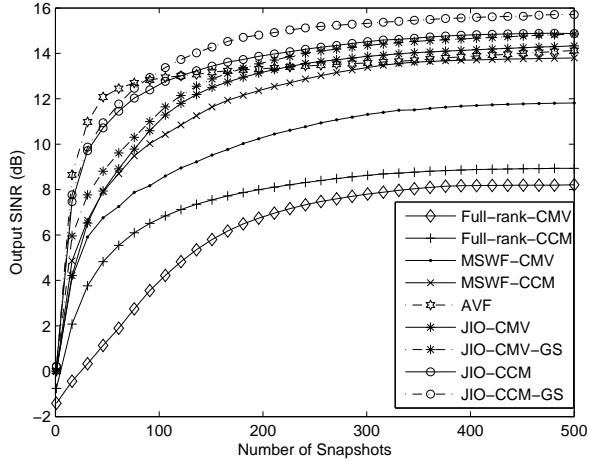,scale=0.6}} \vspace{-1em}\caption{Output SINR versus
dimension ($m$) with $\mu_{T_r}=0.002$, $\mu_{\bar{w}}=0.001$,
$\mu_{T_r,\textrm{gs}}=0.003$, $\mu_{\bar{w},\textrm{gs}}=0.0007$.}
\label{fig:cmv_ccm_sg_gram_rank_final}
\end{minipage}
\end{figure}

In Fig. \ref{fig:cmv_ccm_sg_gram_rank_final}, we keep the same
scenario as that in Fig. \ref{fig:cmv_ccm_sg_gram_final} and check
the rank selection for the existing and proposed algorithms. The
number of snapshots is fixed to $N=500$. The most adequate rank
values for the proposed algorithms are $r=r_{\textrm{gs}}=5$, which
are comparatively lower than most existing algorithms, but reach the
preferable performance. We also checked that these values are rather
insensitive to the number of users in the system, to the number of
sensor elements, and work efficiently for the studied scenarios.

\begin{figure}[htb]
\begin{minipage}[h]{1.0\linewidth}
  \centering
  \centerline{\epsfig{figure=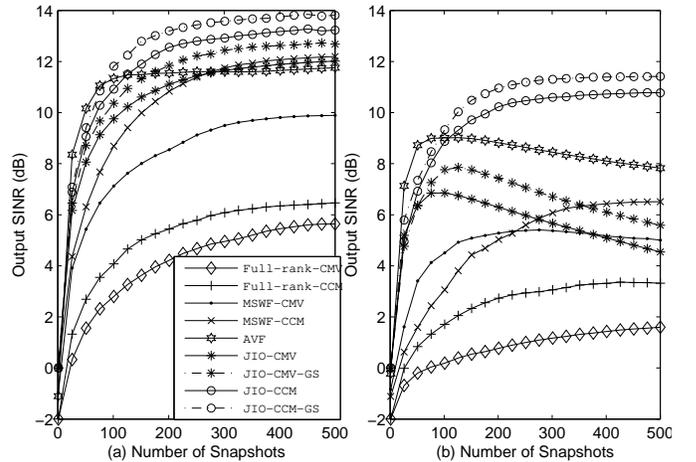,scale=0.6}} \vspace{-1em}\caption{Output SINR
versus the number of snapshots with $\mu_{T_r}=0.002$,
$\mu_{\bar{w}}=0.001$, $\mu_{T_r,\textrm{gs}}=0.003$,
$\mu_{\bar{w},\textrm{gs}}=0.0007$ for (a) ideal steering vector
condition; (b) steering vector mismatch $2^o$.}
\label{fig:cmv_ccm_sg_gram_sve_final_both}
\end{minipage}
\end{figure}

Finally, the mismatch (steering vector error) condition is analyzed
in Fig. \ref{fig:cmv_ccm_sg_gram_sve_final_both}. The number of
users is $q=10$, including one desired user. In Fig.
\ref{fig:cmv_ccm_sg_gram_sve_final_both}(a), the exact DOA of the
SOI is known at the receiver. The output performance of the proposed
algorithms is better than those of the existing methods, and the
convergence is a little slower than that of the AVF algorithm, but
faster than the others. In Fig.
\ref{fig:cmv_ccm_sg_gram_sve_final_both}(b), we set the DOA of the
SOI estimated by the receiver to be $2^{\textit o}$ away from the
actual direction. It indicates that the mismatch problem induces
performance degradation to all the analyzed algorithms. The
CCM-based methods are more robust to this scenario than the
CMV-based ones. The proposed algorithms still retain outstanding
performance compared with other techniques.


\section{Concluding Remarks}

We proposed a CCM reduced-rank scheme based on the joint iterative
optimization of adaptive filters for beamforming and devised two
efficient algorithms, namely, JIO-CCM and JIO-CCM-GS, for
implementation. The transformation matrix and reduced-rank weight
vector are jointly estimated to get the filter output. By using the
GS technique to reformulate the transformation matrix, the
JIO-CCM-GS algorithm achieves faster convergence and better
performance than the JIO-CCM. The devised algorithms, compared with
the existing methods, show preferable performance in the studied
scenarios.

\end{document}